\documentclass[%
 aip,
 jcp,
 amsmath,
 amssymb,
reprint,%
]{revtex4-1}
\usepackage[utf8]{inputenc}
\usepackage[T1]{fontenc}
\usepackage{mathptmx}
\usepackage{graphicx}
\usepackage{dcolumn}
\usepackage{color}
\usepackage{siunitx}
\usepackage{enumitem}
\usepackage{hyperref}
\renewcommand{\v}[1]{\ensuremath{\mathbf{#1}}} 
\newcommand{\gv}[1]{\ensuremath{\mbox{\boldmath$ #1 $}}}

\begin{document}
\title{Hamiltonian and Alias-Free Hybrid Particle-Field Molecular Dynamics}
\author{Sigbj\o rn L\o land Bore} \email[email:
]{s.l.bore@kjemi.uio.no} \affiliation{Department of Chemistry, and
  Hylleraas Centre for Quantum Molecular Sciences, University of Oslo,
  PO Box 1033 Blindern, 0315 Oslo, Norway} \author{Michele Cascella}

\affiliation{Department of Chemistry, and Hylleraas Centre for Quantum
  Molecular Sciences, University of Oslo, PO Box 1033 Blindern, 0315
  Oslo, Norway}
 \begin{abstract}
    Hybrid particle-field molecular dynamics combines standard molecular potentials with density-field models into a computationally efficient methodology that is well-adapted for the study of mesoscale soft matter systems.
    Here, we introduce a new
    formulation based on filtered densities and a particle-mesh formalism that allows for Hamiltonian dynamics and alias-free force computation. This is achieved by introducing a length scale for the particle-field interactions independent of the numerical grid used to represent the density fields, enabling systematic convergence of the forces upon grid refinement. Our scheme generalises the original particle-field molecular dynamics implementations presented in the literature, finding them as limit conditions. The accuracy of this new formulation is 
    benchmarked by considering simple monoatomic systems described by the standard hybrid particle-field potentials. We find that by controlling the time step and grid size, conservation of energy and momenta, as well as disappearance of alias, is obtained.
    Increasing the particle-field interaction length scale permits the use of larger time steps and coarser grids. This promotes the use of multiple time step strategies over the quasi-instantaneous approximation, which is found to not conserve energy and momenta equally well.
    Finally, our investigations of the structural and dynamic properties of simple monoatomic systems show a consistent behavior between the present formulation and Gaussian Core models. 
 \end{abstract}
\date{\today}
\maketitle
\section{Introduction}
Hybrid particle-field simulations (hPF) are a group of computationally
efficient approaches for studying mesoscale soft matter systems with
molecular
resolution.~\cite{Daoulas2006JCP,Muller2011JSP,Milano2009JCP,Vogiatzis2017MACRO}
In hPF models, intermolecular pair interaction potentials, constituting the computationally expensive part of a molecular energy function,
are replaced by particle-field interactions that are functionally dependent on the
densities of the particles composing the system. As a consequence, the
motion of the moieties decouples, yielding a
substantial simplification for the sampling of the phase space. From
an algorithmic point of view, the hPF methods are implemented through particle-mesh (PM) approaches, giving excellent
parallelization efficiency.\cite{Zhao2012JCP} Very recently, a
GPU-based implementation of the Monte Carlo based hPF (\emph{single
  chain in mean field}) (SCMF) set a new milestone with simulations of
polymer melts composed by 10 billion particles.\cite{Schneider2019CPC}

Coupling hPF to molecular dynamics (hPF-MD) algorithms has widened the
range of applicability of hPF systems, from more conventional soft
polymer mixtures to biological
systems.\cite{Milano2013PHYSBIO,Soares2017JPCL,Cascella2015CHEMMOD,Marrink2019CHEMREV} Examples
from the literature include nanocomposites, nanoparticles, percolation
phenomena in carbon
nanotubes,\cite{Nicola2016EPJ,Zhao2016NANOSCALE,Munao2018NANOSCALE,Munao2019MACRO} lamellar and nonlamellar phases of
phospholipids,\cite{Nicolia2012TCA,ledum2020automated,Nicolia2011JCTC}
and more recently polypeptides,\cite{bore2018hybrid}
polyelectrolytes,\cite{Zhu2016PCCP,kolli2018JCTC,Bore2019JCTC,Denicola2020BBA,Schafer2020ANG} and unconventional surfactants.\cite{Carrer2020ChemrXiv}

Given this already outstanding versatility of hPF-MD in modelling soft matter systems, it is 
of great interest to further investigate the intrinsic accuracy of the current methodology, as well as possibilities of extensions that can improve its accuracy and applicability.
hPF-MD models are currently developed such that the grid size determines the range of hPF interactions between particles. If used consistently, this allows for very efficient computation of forces as few grid points are needed in the force computation.\cite{Zhao2012JCP} A disadvantage however, is that this approach does not allow reducing the grid size in order to control the accuracy. Direct dependence of the interactions on the grid size is in particular not ideal for the newly developed constant-pressure simulations methodology,\cite{Bore2020JCP} where the simulation box,  and thus the grid size, changes with time. In this work our goal is to establish a formulation that, being still consistent with the hPF-MD procedure of Milano and Kawakatsu,\cite{Milano2009JCP} additionally allows  for the systematic control of the accuracy 
by grid refinement.

The hPF-MD procedure was first derived using a self-consistent field theory\cite{Milano2009JCP} approach with the forces obtained using a mean field approximation (saddle point approximation). Such approximated forces, as well as hPF potentials commonly displayed with temperature-dependent energy terms, and a terminology of hPF potentials called excess free energies, all suggest that the dynamics of hPF-MD does not conserve the total energy. However, since this pioneering derivation, another approach for the calculation of the forces has been reported by Theodorou and coworkers.\cite{Vogiatzis2017MACRO}
This derivation is not based on statistical mechanics considerations
but on a spatial derivative of the hPF potential energy functional with respect to
the particle position. Despite the methods being derived in two
different manners, the expressions for the forces obtained by the two
methods are very similar, and in fact equivalent under certain implementations, as shown later in this study.
Importantly, the fact that the derivation of particle-field forces in
ref.\cite{Vogiatzis2017MACRO} is based on a standard derivative of
a potential energy with respect to the particle positions, suggests that the hPF-MD procedure is more general than initially thought and that
it is possible to obtain Hamiltonian dynamics that conserves both
the total momentum and energy of the system. 
A Hamiltonian formulation for hPF-MD provides a rigorous basis to investigate the range of validity of simulation parameters, such as the time step, as well as algorithms for time integration such as  the quasi-instantaneous approximation.\cite{Daoulas2006JCP,Milano2009JCP} 

In grid-based numerical procedures, aliasing is recurring problem
that produce unwanted artifacts. In hPF-MD, the most prominent example of reported
aliasing appears when studying molecular assemblies, where
instead of predicting spherical vesicles or droplets, a cubic shape
oriented according to the grid is produced.\cite{Sevink2017soft,
  Bore2020JCP} This particular artifact has two main causes. First, the grid size can be too coarse to represent a spherical shape. This effect is thus more or less relevant according to the characteristic size of the molecular aggregates.\cite{Sgouros2018Macro} Second, the finite difference estimate for the derivative can have a directional bias according to the grid. This can be remedied by the use of a rotational invariant finite
difference scheme.\cite{Sevink2017soft,pizzirusso2017biomembrane,Denicola2020BBA} Additional
aliases, such as breaking of translational invariance,
can possibly produce subtler artifacts, including weak uneven forces eventually hindering the
free diffusion of molecules. Therefore, it is important to develop an alias-free approach that guarantees accurate hPF-MD simulations, and that can also act as a benchmark for other, eventually faster, implementations.

Even though conserving the energy and controlling aliasing appear in principle as two distinct problems, they are in fact related. On the one hand, achieving conservation of energy and momenta in practice requires accurate calculation of the forces. On the other hand, aliasing can effectively be looked upon as a consequence of having insufficient resolution to represent the densities of particles on the scale at that we are interested. Thus, to solve both problems we need a methodology that allows for grid size independent interactions between particles. This necessitates introducing a grid independent length scale controlling the
range of interactions between the particles,  which can be done in two
main ways:  by defining interaction potentials between particles,
such as pair interactions similar as was done in ref.\cite{laradji1994off}, or by defining
interaction energies as function of filtered densities, such as was
done in ref.\cite{Bore2020JCP}, but this time with a filter having an
intrinsic smoothing length scale. For quadratic dependence on densities, these two approaches are largely equivalent, however a filtering approach is more flexible as it allows for any functional dependence on densities. Nevertheless, it remains to rigorously and efficiently compute the corresponding forces. For this purpose, the rich literature of particle-mesh (PM) methods\cite{hockney1988computer,feng2016fastpm} and related implementations are directly applicable without requiring major modification.    

Here, we develop a Hamiltonian and alias-free formulation of hPF-MD 
based on filtered densities, together with its preliminary implementation using PM routines for the force computation, and present benchmarks on simple test systems.

\section{Theory and Methods}
In hPF-MD we consider a system of $N$ particles in $M$ molecules subject to the Hamiltonian:
\begin{equation}\label{eq:H}
    H(\{\mathbf r\})=\sum_{m=1}^M H_0(\{\v r,\dot {\mathbf r}\}_m) + W[\phi(\v r)],
\end{equation}
where $H_0$ is the Hamiltonian of a single non-interacting molecule
$m$, and $W$ is an interaction energy functional dependent on the particle number
densities $\phi$:
\begin{equation}\label{eq:phi}
    \phi(\v r) = \sum^N_{i=1} P(\v r-\v r_i).
\end{equation}
 $P$ is a window function that is used to distribute the particles in the space, in practice done by assigning $P$ onto a grid.  

Sampling of the phase space associated to \eqref{eq:H} using MD requires computing the forces due to both
$H_0$ and $W$. Forces due to $H_0$ are computed using standard
molecular potentials (see ref.~\cite{Rapaport2004CRC}), while forces
due to $W$ require specialized techniques that are at the centre of this work. Therefore, in the
following we will only consider:
\begin{equation}\label{eq:H-mono}
    H=\sum_{n=1}^N\frac 12 m_i\v v_i^2 + W[\phi],
\end{equation}
which amounts to integrating the equations of motion of
monoatomic particles subject to $W[\phi]$.

\subsection{Forces on particles}
The force on particle $i$ due to $W$ is determined by:
\begin{equation}\label{eq:F}
    \mathbf F_i =-\frac{\partial W}{\partial \v r_i} =-\int   \frac{\delta W}{\delta\phi(\v r)}\frac{\partial \phi(\v r)}{\partial \v r_i}\text d\mathbf r,
\end{equation}
where we have employed the chain rule for functional derivatives.
Defining the {\it external potential} $V(\v r)$ as: 
\begin{equation}
  V(\v r)\equiv\frac{\delta W}{\delta \phi(\v r)},
\end{equation}
and inserting \eqref{eq:phi} into \eqref{eq:F}, we obtain:
\begin{equation}\label{eq:F-P}
    \mathbf F_i = -\int   V(\v r)\frac{\partial P(\v r-\v r_i)}{\partial \v r_i}\text d\mathbf r.
\end{equation}
Next, we change the derivative variable:
\begin{equation}\label{eq:F-P2}
    \v F_i = \int   V(\v r)\gv{\nabla} P(\v r-\v r_i)\text d\v r.
\end{equation}
Then, using integration by parts and assuming periodic boundary
conditions, we find:
\begin{equation}\label{eq:F-P3}
    \mathbf F_i = -\int  \gv\nabla V(\v r) P(\v r-\v r_i)\text d\mathbf r.
\end{equation}
We note here that \eqref{eq:F-P3} is a generalization of the original expression reported in ref.\cite{Milano2009JCP} where the forces on the particles are expressed as:
\begin{equation}\label{eq:F-hpfstandard}
    \mathbf F_i = -\gv\nabla V(\v r_i).
\end{equation}
In fact, \eqref{eq:F-hpfstandard} is equivalent to \eqref{eq:F-P3}  when $-\gv\nabla V(\v r_i)$ is computed using the same window function as for the density estimation.  

\subsection{Grid convergent local functionals}
If the window function $P$ is dependent on the grid size, as it is customary in PM
methods to avoid very expensive density and force interpolation, then  the
density $\phi(\v r)$ is dependent on the grid size, and consequently also $W$. This situation corresponds to the standard hPF-MD\cite{Milano2009JCP} and SCMF. To obtain a methodology without such grid biasing, we employ here a filtered density:
\begin{equation}
    W= W[\tilde\phi([\phi])],\quad \tilde \phi(\v r)\equiv \int \phi(\v x)H(\v r-\v x)\text d \v x,
\end{equation}
where $H$ is a grid independent window function that smooths out $\phi$,
and ensures that both $\tilde\phi$ and $W$ converge as the grid size is reduced.

The external potential acting on a particle is then given by:
\begin{equation}
 V(\v r)=\int\frac{\delta W}{\delta \tilde \phi(\v y)}\frac{\delta \tilde \phi(\v y)}{\delta \phi(\v r)}\text d\v y.
\end{equation}
Assuming local dependency of $W$ on $\tilde \phi$:
\begin{equation}
    W[\tilde\phi]=\int w(\tilde\phi(\v r))\text d\v r,
\end{equation}
and using:
\begin{equation}
    \frac{\delta \tilde \phi(\v y)}{\delta \phi(\v r)}=H(\v y-\v r),
\end{equation}
we find:
\begin{equation}
 V(\v r)=\int\frac{\partial w}{\partial \tilde \phi}(\v y)H(\v y-\v r)\text d\v y.
\end{equation}

\subsection{hPF-MD and PM}
Force computation by a PM approach can mainly be done in two ways,
either by derivative of the assignment function through \eqref{eq:F-P},\cite{Vogiatzis2017MACRO} or
by interpolating the derivative of the external potential.\cite{Deserno1998JCP} Here we choose to
interpolate the derivative of the external potential. The PM approach
consists of the following three main steps.
\subsubsection{Computation of density on a grid}
The estimation of discrete densities requires specifying the window
function $P$. The three most important window functions proposed in the literature are
Nearest-Grid-Point (NGP), Cloud-In-Cell (CIC) and Triangular-Shaped
Cloud (TSC), which differ by considering one, two and three grid
points per dimension respectively.\cite{hockney1988computer} In the
following we only consider CIC. The density is computed at grid point $ijk$
point by:
\begin{equation}
    \phi_{ijk}=\sum^N_{k=1}P(\v r_{ijk}-\v r_k).
\end{equation}

\subsubsection{Determination of the external potential} 
Considering functionals locally dependent on $\tilde\phi$, the first step is to obtain $\tilde \phi(\v r)$. A straightforward way of
obtaining it is by \textit{Fast Fourier Transform} (FFT):
\begin{equation}
    \tilde \phi_{ijk}=\text{FFT}^{-1}\left[\text{FFT}(\phi)\text{FFT}(H)\right],
\end{equation}
where we have used that a convolution is a product in
Fourier space. Next, we find the external potential as:
\begin{equation}
    V_{ijk}=\text{FFT}^{-1}\left[\text{FFT}\left(\frac{\partial w(\tilde\phi(\v r))}{\partial \tilde\phi}\right)\text{FFT}(H)\right].
\end{equation}
The derivative of $V$ can be obtained by finite differences, but for
smooth $V$, it is best computed by:
\begin{equation}
    \gv\nabla V_{ijk}=\text{FFT}^{-1}\left[i\v k~\text{FFT}\left(\frac{\partial w(\tilde\phi(\v r))}{\partial \tilde\phi}\right)\text{FFT}(H)\right].
\end{equation}
\subsubsection{Force interpolation}
The forces are computed by interpolating back the derivative of the
external potential onto the particles through
\eqref{eq:F-P3}:
\begin{equation}\label{eq:F-P4}
    \mathbf F_i = -\sum_k\gv\nabla V_{j_k} P(\v r_{j_k}-\v r_j)h^3,
\end{equation}
where $j_k$ is the neighbouring vertices of particle $j$ and $h^3$ is
the volume of a single cell. We note that in the limit of a very small
grid size $h$, the force will be independent of the choice of $P$, and it is
possible to use $P'$ for the force interpolation. However, for a finite grid, using $P'$
can lead to artifacts, while having the same $P$ ensures conservation of both the 
momenta and the energy.\cite{hockney1988computer}

\subsection{Hybrid particle-field model}
As a model for intermolecular interactions we consider the standard energy mixing potential commonly adopted in hPF-MD\cite{Milano2009JCP} and SCMF,\cite{Daoulas2006JCP}  this time defined using filtered densities:
\begin{equation}\label{eq:W-FH}\footnotesize{
    W[\tilde\phi]=\frac 1{\phi_0}\int\left(\sum_{k<\ell}\tilde\chi_{k\ell}\tilde\phi_k(\v r)\tilde\phi_\ell(\v r)+\frac 1{2\kappa}\left(\sum_\ell\tilde\phi_\ell(\v r)-\phi_0\right)^2\right)\text d \v r,}
\end{equation}
where $\tilde\chi_{k\ell}$ is the Flory-Huggins mixing parameter between particle species $k$ and $\ell$, $\kappa$ is a
compressibility parameter and $\phi_0$ is the average density of the system. The
corresponding external potential is given by:
\begin{equation}\small{
    V_k(\v r)=\frac 1{\phi_0}\int\sum_{\ell}\left(\tilde\chi_{k\ell}\tilde\phi_\ell(\v x)+\frac 1\kappa\left(\sum_\ell\tilde\phi_\ell(\v x)-\phi_0\right)\right)H(\v x-\v r)\text d \v x.}
\end{equation}
The full specification of the model requires defining $H$, the grid
independent window function. The fields of \emph{spectral
  methods}\cite{canuto2006spectral} and \emph{Kernel density
  estimation}\cite{parzen1962estimation} offer several choices for $H$. Here, for simplicity, we use the Gaussian:
\begin{equation}
    H(x)=\frac{1}{\sqrt{2\pi}\sigma}e^{-\frac {x^2}{2\sigma^2}},
\end{equation}
and accordingly its Fourier Transform:
\begin{equation}
    \hat H(k)=e^{-\frac 12\sigma^2k^2},
\end{equation}
where the standard deviation $\sigma$ is an indication of the space occupied by the
particle. 
We remark that the choice of a Gaussian filter for the case of quadratic density-dependent interaction potentials effectively corresponds to the Gaussian Core model (see Appendix~\ref{app:gaussian}).\cite{stillinger1997negative,pike2009theoretically,laradji1994off} However, the derivation of the particle forces is general beyond such quadratic terms, and the methodology is applicable for any functional local or nonlocal dependency on density.  

\subsection{Implementation and simulation parameters}

\subsubsection{Implementation of hPF-MD}
The hPF-MD routines as described above for monoatomic particles were
implemented in a simple Python code. Besides basic implementations of
velocity-Verlet integrator\cite{swope1982computer} and thermostat by
velocity-rescaling,\cite{Bussi2007JCP} it contains PM routines for
force calculation which are based on the pmesh
package\cite{yu_feng_2017_1051254}. The full hPF-MD code used to produce all the results contained in the result and discussion section
is freely available on \url{https://github.com/sigbjobo/hPF\_MD\_PMESH}.

All the details for the systems simulated  in this study are provided in Appendix~\ref{appendix:sim-setup}. 
\subsubsection{Dimensional analysis}
\begin{table}[!htb]
\caption{Conversion factors for dimensionless units.}\label{tab:dimensionless}
\begin{tabular}{c|c|c|c|c|c|c}
\hline\hline
    $l_0$&$E_0$&$t_0$&$v_0$&$m_0$&$ F_0$&$k_{\text b}T_0$\\
    \hline
    ${\phi_0}^{-1/3}$&$\kappa^{-1}$&$\phi_0^{1/3}\sqrt{\kappa^{-1}/m}$&$\sqrt{\kappa^{-1}/m}$&$m$&$\kappa^{-1}\phi_0^{1/3}$&$\kappa^{-1}$\\
\hline\hline
\end{tabular}
\end{table}
The simplicity of the model potential \eqref{eq:W-FH} allows us to
formulate a set of dimensionless units that reduce
considerably the parameter space. Let a quantity $a$ and its
dimensionless counterpart $a^*$ be related by:
\begin{equation}
    a=a_0a^*,
\end{equation}
then $a_0$ is the conversion factor. The conversion factors are
reported in Table~ \ref{tab:dimensionless}. Since $\phi_0$, $\kappa$
and $m$ just appear as conversion factors, these variables can be put
constant throughout all the simulations. In particular, this corresponds to that lengths are measured in terms of the specific length of each particle and that the effect of the compressibility parameter is determined by the overall temperature/energy of the system. In summary, to probe properties of this model our simulations need only to consider
simulation settings grid size $h^*$, time step $\Delta t^*$, model
properties $\sigma^*$ and $\chi^*$, under thermodynamic conditions $E^*$ or
$T^*$, and $N$. 

\section{Results and Discussion}
\subsection{Conservation of momenta and energy}
\begin{figure*}[t]
\includegraphics{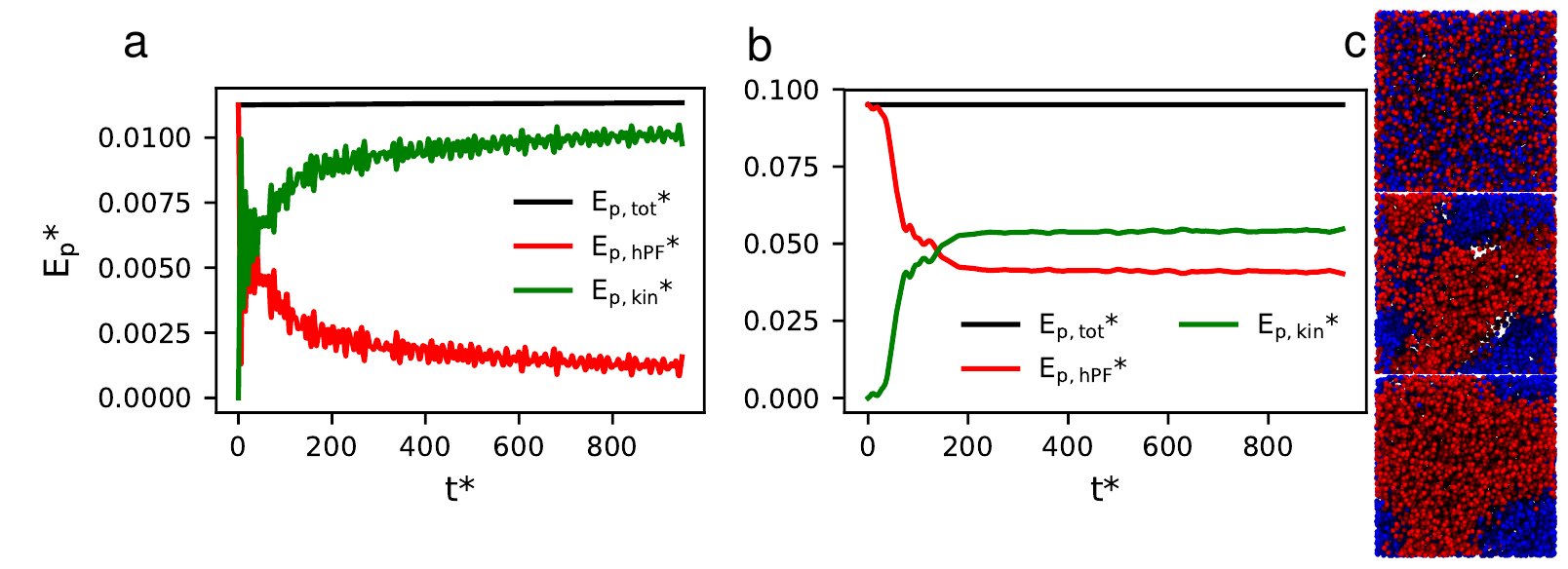}  
\caption{Energy per particle ($E_{\text p}^*$ as function of time for a system of initially randomly dispersed particles with zero velocity. {\it Panel a:} single component system; {\it panel b:} binary, phase-separating system; {\it panel c:} Representative snapshots along the time evolution of the binary system.
}\label{fig:conservation-energy}
\end{figure*}
\begin{figure}[!htb]
\includegraphics{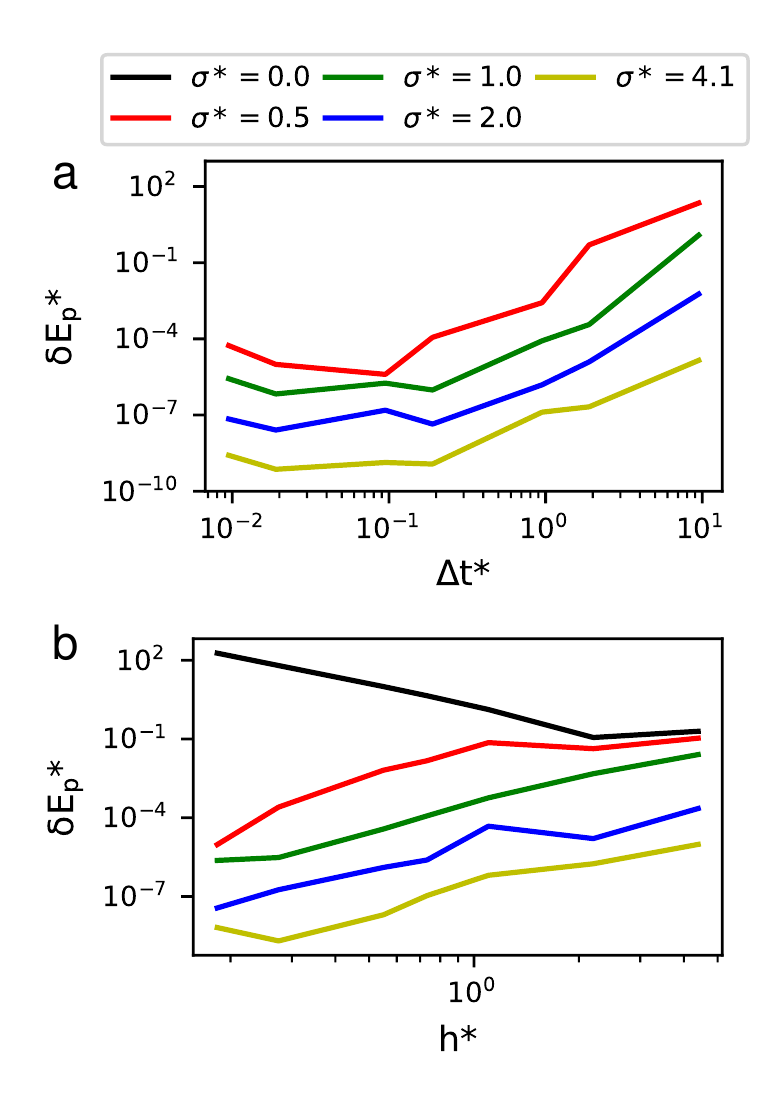}  
\caption{Change in total energy per particle as function of time step $\Delta t^*$ ({\it panel a}),  and grid size $h^*$ ({\it panel b}), for different values of $\sigma^*$.}\label{fig:E-dt-h}
\end{figure}
As an initial test for the ability of the proposed formulation of hPF-MD to
conserve energy we consider a system of randomly dispersed particles
with initial zero velocity, simulated under $NVE$ conditions. As seen in
Figure~\ref{fig:conservation-energy}, the system quickly heats up as the
potential energy is exchanged with the kinetic energy,
while the total energy remains conserved. As expected by using
the same interpolation for density and forces,\cite{hockney1988computer} the total momentum
remains identical to zero throughout the whole simulation (data not shown). 
The conservation of the energy is perfectly observed also when simulating a binary mixture of repulsive particles, where dramatic structural changes like the progressive phase separation of the two components do not affect the total energy of the system  (Figure~\ref{fig:conservation-energy}b,c).

The degree of conservation of energy, as with any molecular
simulation, will generally depend on the total energy of the molecular
system, but here also on the grid size $h^*$, time step
$\Delta t^*$ and $\sigma^*$. Figure~\ref{fig:E-dt-h}
reports how the conservation of energy depends on these
parameters. Using a small grid size $h^*$, we can control for the bias due to grid
discretization and examine the effect of reducing the time step (Figure~\ref{fig:E-dt-h}a). As
expected, reducing the time step improves energy
conservation. Interestingly, the $\sigma^*$ parameter has two important
outcomes. First, conservation of energy is improved by increasing
$\sigma^*$. Second, the plateau, at which decreasing  $\Delta t^*$ does not  improve the energy conservation,  is reached for larger time steps. This agrees
with the general trend of $\sigma^*$ being scale of coarse-graining leading to smoother
dynamics, and thereby allowing for larger time steps. 

Running simulations using
a very small time step and examining
the conservation of energy as function of $h^*$, we generally observe
better convergence by refining the grid (Figure~\ref{fig:E-dt-h}b). However, for grids spacing $h^*$ larger 
than 1, corresponding to the specific length between particles, we also obtain a good conservation of energy, even without
smoothing. This regime corresponds to standard hPF-MD in which density
distribution itself is sufficiently smooth to achieve conservation of the energy. We note if it is desirable to reduce the smoothing obtained by increasing the grid size, one possible solution is to apply an additional sharpening\cite{feng2016fastpm,hockney1988computer} on the unfiltered densities $\phi$. 

\subsection{Aliasing}
\begin{figure}[!htb]
\includegraphics{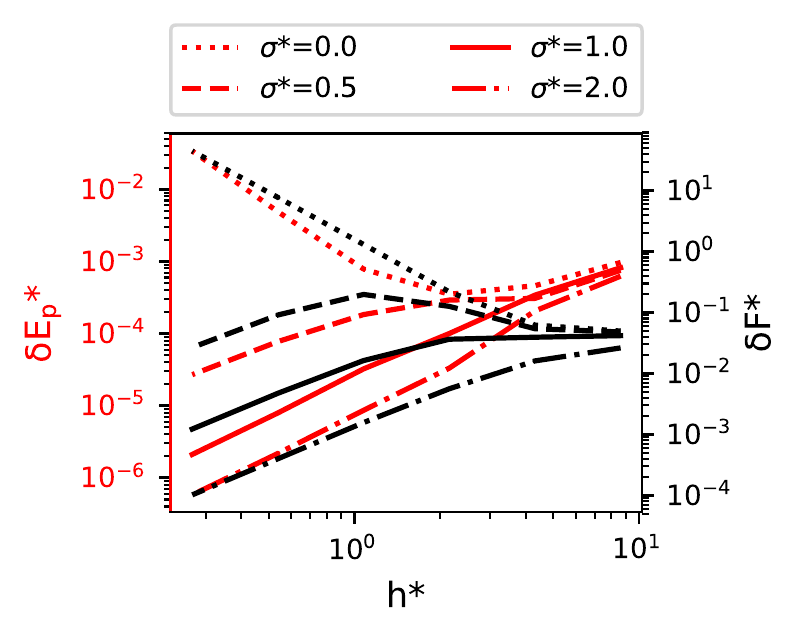}
\caption{The standard deviation of the energy (red line and axis) and the forces (black lines and axis) averaged over
  100 random translations and rotations for a system of \num{10000} randomly
  dispersed particles. }\label{fig:trans-rot-FE}
\end{figure}
To evaluate aliasing due to the grid on the force calculation, we
consider the energy and the forces in a reference system of randomly
dispersed particles before and after a random rototranslation operation (see Appendix~\ref{appendix:sim-setup} for more details). We
measure the effect of the grid by computing standard deviations of
$\delta E^*$ and $\delta F^*$ (Figure~\ref{fig:trans-rot-FE}). The general
trend observed for conservation of the energy as function of $h^*$ is also
mirrored in Figure~\ref{fig:trans-rot-FE}. Reducing $h^*$ reduces
aliasing, unless for large $h^*>1$, at which the smoothing due to density
interpolation itself reduce aliasing. Thus, to avoid
significant aliasing, the grid either needs to be smaller than
$\sigma^*$ or bigger than 1, the average distance between particles.

\begin{figure}[!htb]
\includegraphics{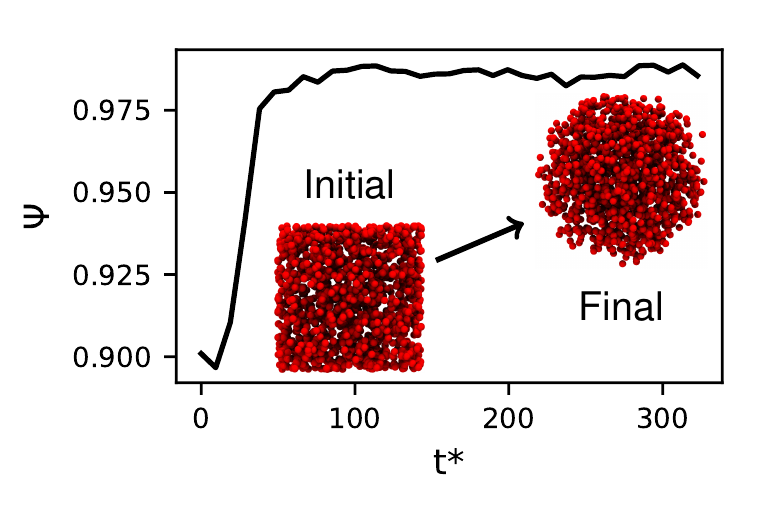}
\caption{Time evolution of the sphericity of a  droplet of a monoatomic fluid in an imiscible solvent. Snapshots depict the initial organization of the droplet prepared in a cubic shape, and the final spherical organization found by hPF-MD. The solvent phase is not displayed for clarity purposes.}\label{fig:sphericity}
\end{figure}
Aliasing in hPF-MD has been particular noticeable when considering molecular aggregates, such as in large vesicles or in droplets, where instead of a spherical shape, a cubic shape oriented according to the grid has been predicted. There are primarily two causes: aliasing due to insufficient grid resolution to represent spherical shapes (such aliases are cured when considering larger systems\cite{Vogiatzis2017MACRO}), and numerical derivatives of the external potential that are not rotational invariant\cite{Sevink2017soft,Denicola2020BBA}. Here, since we can arbitrarily reduce the grid size, and the derivative is estimated  in Fourier space, both of these forms of aliasing can be controlled for. To demonstrate this, we here consider the ability of our formulation to predict the sphericity $\Psi$ of a phase-separated droplet:
\begin{equation}
    \Psi=\frac{\pi^{1/3}(6V)^{2/3}}{A},
\end{equation}
where $V$ and $A$ are the volume and the area of the droplet. Figure~\ref{fig:sphericity} reports the time evolution of the sphericity of the droplet from a prepared cubic shape as function of time. As seen, the sphericity quickly increases from $0.90$ to $0.99$, which is close to a perfectly spherical object.

\subsection{Quasi-instantaneous approximation}
\begin{figure}[!htb]
\includegraphics{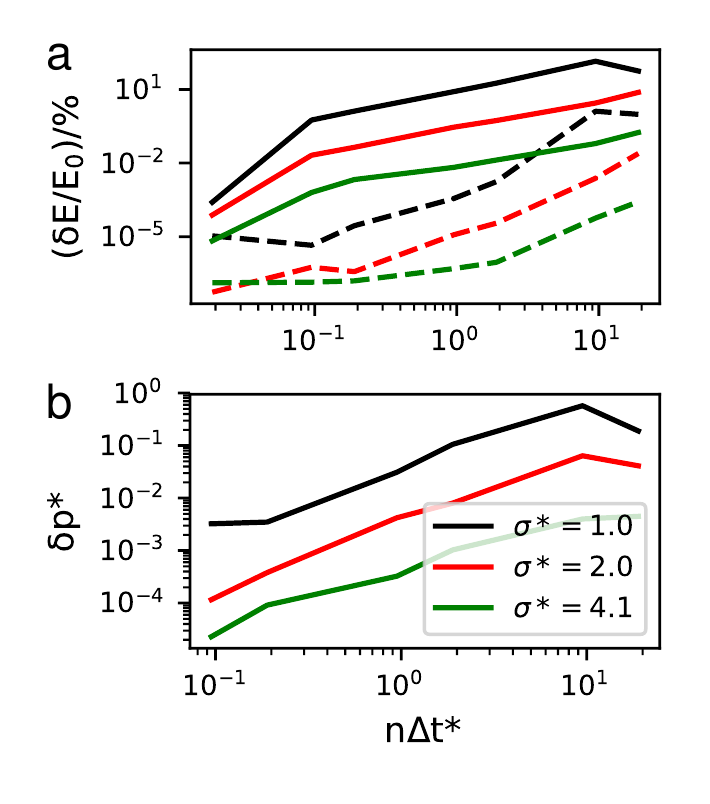}
\caption{Comparison between quasi-instantaneous approximation (whole lines) integrating $n$ times the equations of motion with time interval $\Delta t^*$, and direct integration with a time step $n\Delta t^*$ (dashed lines). {\it Panel a:} Relative deviation of the total energy. {\it Panel b:} Change in
  average momentum for the quasi-instantaneous approximation. No change in momentum was found using large integration steps.
}\label{fig:quasi2}
\end{figure}
In SCMF and hPF-MD, the quasi-instantaneous approximation is used to speed up simulations by approximating the external potential as constant between multiple time steps.\cite{Daoulas2006JCP,Milano2009JCP} For hPF-MD in particular, it can be defined as:
\begin{align}
     \mathbf F^{\text{quasi}}_i(t,t')&= -\int  \gv\nabla V(\v r,t) P(\v r-\v r_i,t')\text d\mathbf r,
\end{align}
where $t$ is the time of the external potential update, is $t'$ is the
time of the force computation ($t\le t'$).  A direct measurement of how well the
quasi-instantaneous approximation works is obtained by running
simulations and recording change in energy and momenta after a certain
amount time (Figure~\ref{fig:quasi2}). The
quasi-instantaneous approximation not only produces a net momentum, but also achieves worse conservation of energy than simply using a
larger time steps. This is surprising, as it has been demonstrated that such approximation achieves good results in SCMF.\cite{Daoulas2006JCP} However, contrary to SCMF, hPF-MD approximates forces and not changes in the energy as required by Monte Carlo. While perhaps disappointing, we remark that this approximation is likely to be more relevant for SCMF, which requires a stationary external potential between Monte Carlo moves in order to update multiple particle positions in between external potential updates.
The same approximation may not be as much needed in hPF-MD. In this case, when large time steps may be employed, depending on the value of $\sigma^*$, a multiple time step algorithm in which particle-field forces are kept constant in between external potential updates and intramolecular potential forces are computed frequently is a promising alternative.\cite{tuckerman1992reversible} 

\subsection{Properties}
\begin{figure}[!htb]
   \includegraphics{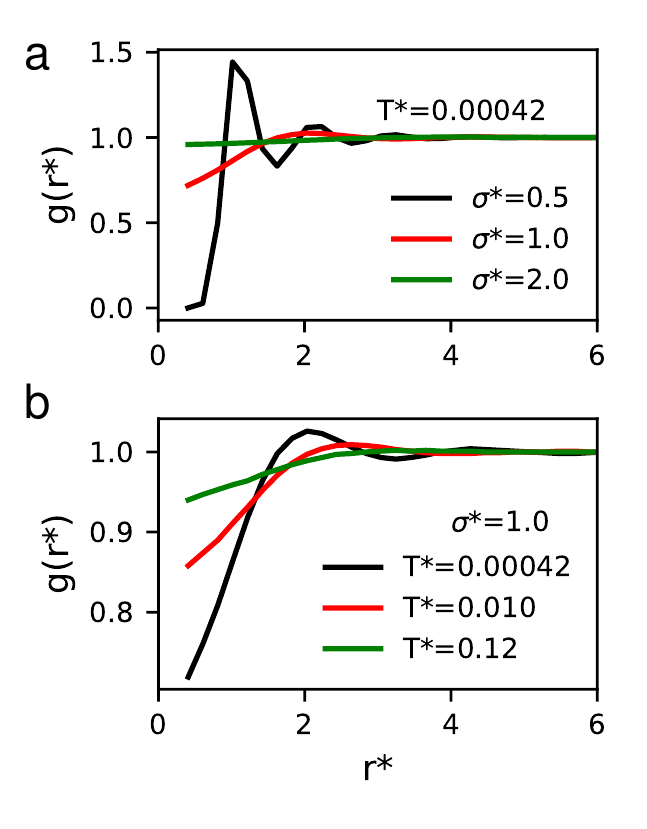}
  \caption{ Radial distribution function ($g(r^*)$) for a single component monoatomic fluid. {\it Panel a:} Low temperature behavior for different values of $\sigma^*$. {\it Panel b:} Temperature dependence of $g(r^*)$ for $\sigma^* = 1$.} \label{fig:rdf}
\end{figure}
Particle-field interactions are designed to enforce thermodynamic averages, such as partitioning and local density fluctuations. Nevertheless, the presence of a well defined interaction potential will necessarily introduce some spatial organisation among the molecules composing the system. In Figure~\ref{fig:rdf} we report the radial distribution function for a monoatomic fluid simulated at $NVT$ conditions. We first note that the lower the temperature, the more structure is observed, with the appearance of a solvation shell structure similar to that of a liquid phase. Inspecting the radial distribution function at very low $T^*$, which according to our reduced units corresponds to a low level of compressibility, we examine the effect of the smoothing parameter $\sigma^*$ on the structure of the fluid. Overall, by lowering $\sigma^*$ more structuring is obtained. 
This effect can be interpreted by considering $\sigma^*$ as a measure of the delocalization of the particles. In the limit of $\sigma^*\gg 1$, the particles are delocalized in the whole volume; thus, any conformation would produce the same local density equal to the average density, and very little structuring of the fluid. On the contrary, for $\sigma^*\sim 0.5$ the particle spread corresponds to the specific length; in this case, an organisation of the particles at a distance $r*~1$ will guarantee a homogeneous local density. Instead, any motion will produce  fluctuations of the density, and a consequent increase in the energy. As a consequence, the hPF induces strong structuring in the fluid at the characteristic $r^*=1$ distance (Figure~\ref{fig:rdf}). For lower values of $\sigma^*$, particles will be localised more than the range of the external potential, thus the model will produce a behavior similar to that of a diluted gas of small hard spheres. This overall behaviour is consistent with what has been reported for the Gaussian Core model\cite{stillinger1997negative,laradji1994off}, indicating a direct correspondence between these two approaches. 

\begin{figure}[!htb]
\includegraphics{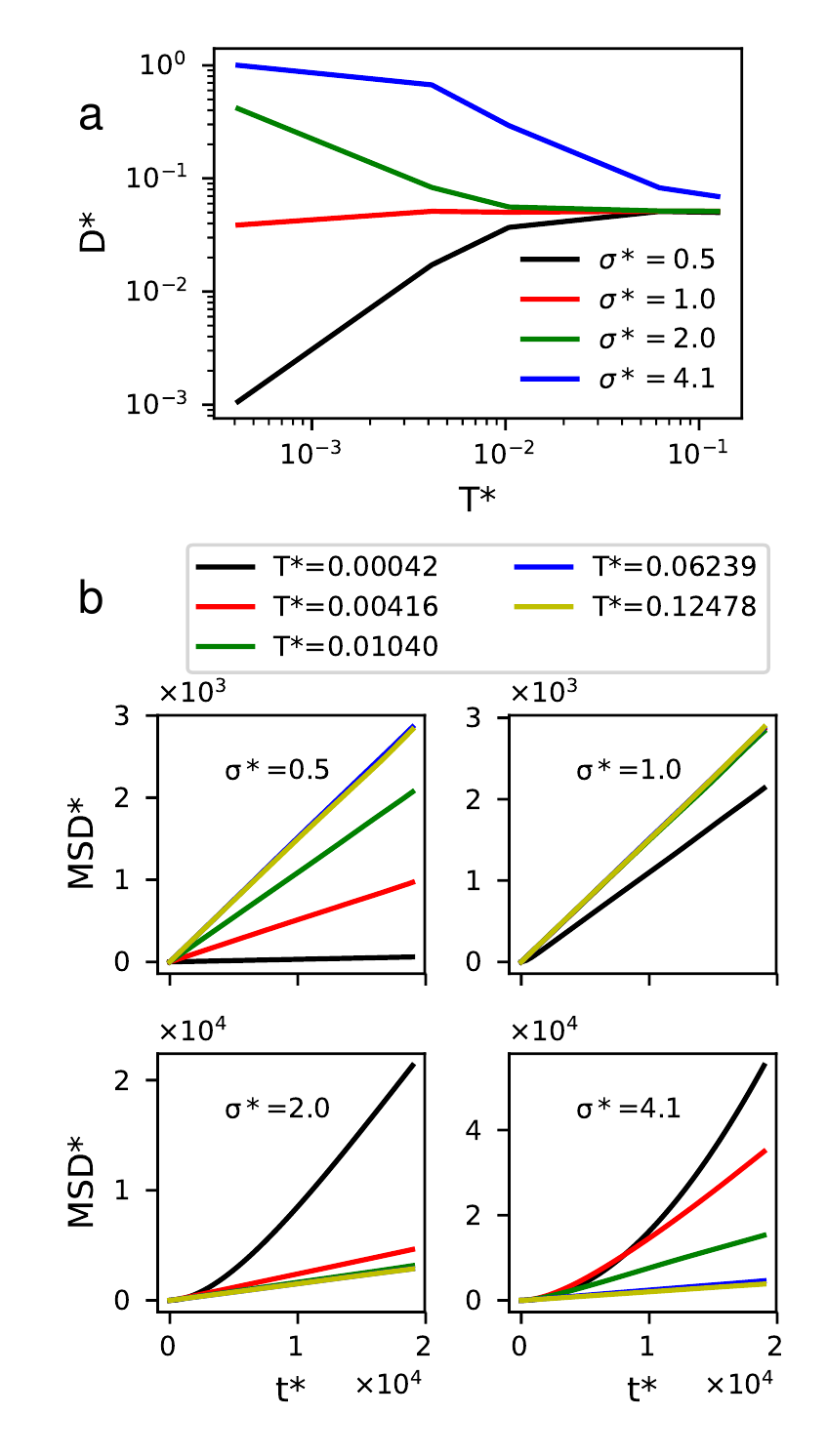}
\caption{{\it Panel a:} Fitted diffusion coefficient ($D^*$) for a single component monoatomic fluid as function of $T^*$ for different values of 
  $\sigma^*$. {\it Panel b}  corresponding Mean square diplacement (MSD) used to compute the diffusion coefficients.}\label{fig:dif}
\end{figure}
Using an appropriate Canonical sampling thermostat,\cite{Bussi2007JCP} we investigate the dynamic behavior of the model. In Figure~\ref{fig:dif} we report both the estimated diffusion coefficient and the mean square deviations for the monophase system at different temperatures.  Increasing $\sigma^*$ leads to higher diffusion, in line with the results reported by Stillinger\cite{stillinger1978study}. This is consistent with interactions being smoother, and $\sigma^*$ corresponding to a scale of coarse-graining, thereby allowing particles to more easily go unhindered. Despite the simplicity of the model, the diffusion behavior as function of temperature is complex. Contrary to the intuitive increase of diffusion with temperature, which corresponds directly to the thermal velocities of the particles, we here observe here for $\sigma^*>1$ a decrease of the diffusion coefficient and for $\sigma^*<1$ an increase. 
A possible explanation of this behavior is that for large $\sigma^*>1$ and low temperatures, the motion of the particles becomes close to ballistic, and as temperature is increased the increased thermal velocity is not compensated by the interaction with density fluctuations that hinder the free diffusion of the particles. For low $\sigma^*<1$ particles will not move in a smooth density landscape and thus this effect is not present and we have the expected increase in diffusion coefficient with temperature. We note that anomalous diffusion behavior has been reported for Gaussian Core models\cite{Mausbach2006FPE}.
From a pragmatic point of view, where diffusion is often the limiting step to achieve good equilibration of the systems under study, increasing the $\sigma^*$ parameter may be an excellent way of speeding up the dynamics of the system and the overall sampling, and thus to reach preequilibrated configurations in relatively short time. 

\begin{figure}[!htb]
  \includegraphics{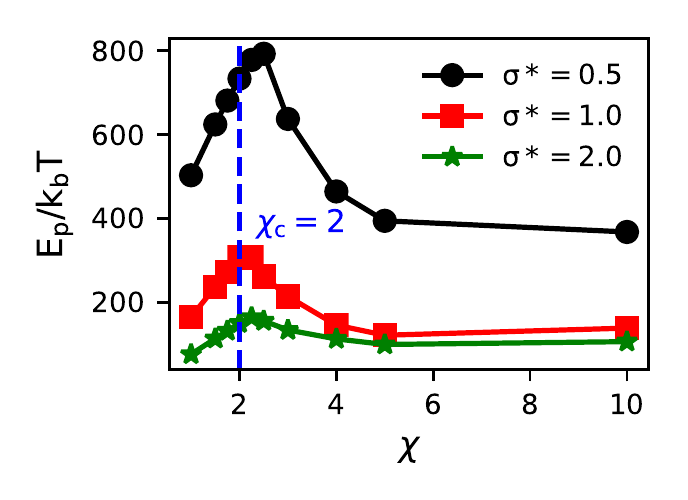}
\caption{Total energy as a function of $\chi=\tilde\chi/k_{\text b}T$ for a fluid composed by two repulsive monoatomic species using different $\sigma^*$ values. The blue dashed lines reports the theoretical mean field critical value $\chi_c$ for the phase separation.  }\label{fig:chi-Echi}
\end{figure}
Finally, in Figure~\ref{fig:chi-Echi} we probe how the $\chi$ parameter
controls the phase behavior of the binary system by computing the
total energy per particle as a function of $\chi$. For all
$\sigma^*$s, we find a breaking point for the first derivative at around $\chi=2.25$,
signaling a phase transition. This value corresponds fairly well to the Flory-Huggins mean field critical point ($\chi=2$). A small overestimation has been previously reported by other hPF approaches\cite{pike2009theoretically} and is likely due to finite size  effects, as well as our simulations not being mean field. The correct representation of the phase separation thermodynamics for different values of $\sigma^*$ is a direct indication that the MD scheme here introduced describes the correct partitioning physics by the hPF model. 

\section{Conclusions}
In this work we have developed a new methodology for achieving
Hamiltonian and alias-free hPF-MD. This development is made possible by three main
steps. First, the hPF-MD forces is derived by spatial
derivative of the interaction energy functional with respect to
particle positions. Second, the hPF-MD functional is expressed in
terms of filtered densities that converge as the grid is
refined. This achieves smooth forces, as well as controlling for
aliases. Last, a standard PM FFT-based implementation for the force
calculation is used to ensure that the theoretical energy and
momentum conservation is achieved. 

Through a series of benchmark studies on simple monoatomic species, we demonstrate  
that the energy and momentum of the particles
in the model are both numerically conserved. The conservation of the energy is primarily
determined by the appropriate choice for the grid size, time step and spread $\sigma$ used for the filtered densities. In accordance with considering $\sigma$
as a scale of coarse-graining, we find that the larger the spread the
larger time steps can be used. If a large enough grid size is
used, larger than the typical distance between particles, well behaved
forces are also obtained without additional smoothing.

The quasi-instantaneous approximation was examined in detail and
benchmarked against using a large time step. We find that this approximation yields worse
conservation of the energy, and that it leads to production of nonphysical net
momenta. This suggests that while the quasi-instantaneous approximation may work well
for Monte Carlo schemes, the same scheme is not ideal for hPF-MD. From an algorithmic point
of view, this suggests that a multiple step algorithm with
frequent computation of intramolecular forces and infrequent
calculation of particle-field forces is a promising alternative.
Such an approach would also not require storing the external potential, as it is needed with quasi-instantaneous approximation, and thereby simplifying parallelization and decreasing memory use of current implementations.\cite{Zhao2012JCP} 

By studying the properties of model, focusing on the effect of Gaussian spread $\sigma$, we find a good agreement between this implementation of hPF-MD and the literature of Gaussian core models. Generally, we find that increasing the spread of the Gaussian reduces the molecular structure and increases the diffusion of the particles.

In summary, we have presented a new formulation of hPF-MD that guarantees numerical robustness and controlled convergence for energy and force calculations. The natural continuation of this work lies in developing an efficient parallelized implementation that also includes molecular potentials. 
\section{Acknowledgements}
Authors acknowledge the support of the Norwegian Research
Council through the CoE Hylleraas Centre for Quantum
Molecular Sciences (Grant No. 262695) and the Norwegian
Supercomputing Program (NOTUR) (Grant No. NN4654K). MC acknowledges funding by the
Deutsche Forschungsgemeinschaft (DFG) within the project B5 of the TRR 146 (project number
233630050). Lastly, the authors would like to acknowledge Yu Feng for developing and distributing the PMESH python package, which allowed for a straightforward implementation of the methodology presented in this study.  

\section*{Data Availability Statement}
 The data that support the findings of this study are available from the corresponding author upon reasonable request.

\appendix
\section{Equivalence with Gaussian-core model}\label{app:gaussian}
For simplicity we here consider a single species:
\begin{align}
    W[\tilde\phi]&=\frac{1}{2\kappa\phi_0}\int\tilde \phi(\v r)\tilde\phi(\v r)\text d \v r
\end{align}
From Plancherel theorem we can rewrite this energy as:
\begin{equation}
    W[\tilde\phi]=\frac{1}{2\kappa\phi_0}\int\hat{\tilde \phi}(\v k)\hat{\tilde \phi}(\v k) \text d \v k.
\end{equation}
Next, inserting for $H$ we find:
\begin{equation}
    W[\tilde\phi]=\frac{1}{2\kappa\phi_0}\int\hat H(\v k)^2\hat{\phi}(\v k)\hat{\phi}(\v k) \text d \v k.
\end{equation}
Next, considering a general pair interaction:
\begin{equation}
    W=\frac {1}{2\kappa\phi_0}\int\phi(\v x)\phi(\v y) K(\v x-\v y) \text d \v x\text d \v y,
\end{equation}
and rewriting as follows:
\begin{equation}
    W=\frac {1}{2\kappa\phi_0}\int\left(\int \text d \v x\phi(\v x)K(\v x-\v y)\right)\phi(\v y)  \text d \v x,
\end{equation}
we again use Plancherel theorem and the convolution theorem for the integral with respect to $\v x$:
\begin{equation}
    W=\frac{1}{2\kappa\phi_0}\int \hat{K}(\v k)\hat \phi(\v k)\hat \phi(\v k).
\end{equation}
We identify here that:
\begin{equation}
    \hat K(\v k)= \hat H^2(\v k).
\end{equation}
For a Gaussian core model we have that:
\begin{equation}
    K(\v k)=e^{-\sigma^2 k^2}
\end{equation}
Thus the square of the filter corresponds to a Gaussian core model.
\section{Simulation setups}\label{appendix:sim-setup}
This section summarize all the simulation settings used to produce all the figures contained in this manuscript.
\paragraph*{Figure 1}
Both graphs were obtained with $NVE$ simulations of a initial system of 10000 particles randomly placed inside a box of volume $L^*=21.54$. A time step $\Delta t^*=0.0019$ and grid spacing $h^*=0.36$ was used. Figure~\ref{fig:conservation-energy}a has only a single particle type, and employed $\sigma^*=1$. Figure~\ref{fig:conservation-energy}b,c considered a $\SI{50}{\percent}/\SI{50}{\percent}$ mixture of the two species, and employed $\sigma^*=2$ and $\tilde\chi^*=0.375$.
\paragraph*{Figure 2}
The system consists of 10648 particles placed centered cubic, in a box of length $L^*=23$. Particles are distributed with an velocity distribution corresponding to a Boltzmann distribution with a temperature of $T^*=0.125$ and simulations last $t^*=94.87$. Figure~2a) employs a $\Delta t^*=0.00379$, while Figure~2b) employs a $h^*=0.183$.
\paragraph*{Figure 3}
This figure considers a box extending $2L^*$ where a system 10000 particles randomly placed inside cubic cell extending $L^*=21.54$. The particles are rotated first by a uniformly distributed rotation and then displaced by a random translation. Rotating back the force, the standard deviation of the forces and the energy is computed over 100 samples. We remark that because of periodic boundary conditions, there is not a rotational invariance, but since we here consider a large box and only have short range interactions, we can do the benchmark as described.
\paragraph*{Figure 4}
The simulations consider 10000 particles in a box of $L^*=21.54$ with a 1500 particles forming as a cube. A $\sigma^*=1$ used and $\chi=10$ is used. A time step of $\Delta t^*=0.019$, $h^*=0.36$ is used. The temperature is controlled by a thermostat with time coupling constant $\tau^*=1.9$ at a temperature of $T^*=0.010$. The area and the volume of the phase-separated shape, which is used for the sphericity, is calculated using the marching cubes algorithm~\cite{lorensen1987marching}.    
\paragraph*{Figure 5}
This figure was made considering a system of 10648 particles placed centered cubic in a box of length $L^*=23$, with initial velocities corresponding to a Boltzmann distribution of temperature $T^*=0.12$. The difference in energy for a $NVE$ simulation is computed after simulations lasting a total time of $t^*=19$.
\paragraph*{Figure 6 and 7}
Both figures consider the same system of 10000 particle of single a type inside a volume $L^*=21.54$ for a simulation lasting $t^*=18973$ with $h^*=0.36$ $\Delta t^*=0.095$.   
\paragraph*{Figure 8}
The binary system is composed of 80000 particles (50\%/50\% mixture) with a box size of $L^*=43$, thermostated at $T^*=0.124$.. Data is gathered for simulations lasting $t^*=18973$ using time step $\Delta t^*=0.19$ and grid size $h^*=1.09$.   
\section*{References}
\bibliography{mybib}
\end{document}